\documentclass[journal]{IEEEtran}

\ifCLASSINFOpdf
\else
   \usepackage[dvips]{graphicx}
\fi
\usepackage{url}

\usepackage{graphicx}
\usepackage{amsmath}
\usepackage{amsfonts}
\usepackage{algorithm}
\usepackage{algorithmic}
\usepackage{xcolor}
\newcommand{\rev}{\textcolor{black}}
\newcommand{\rrev}{\textcolor{black}}
\usepackage{hyperref}
\begin{document}

\title{\rev{Hybrid Liquid Neural Network-Random Finite Set Filtering for Robust Maneuvering Object Tracking}}

\author{Minti Liu, Qinghua Guo, \IEEEmembership{Senior Member, IEEE}, Cao Zeng, \IEEEmembership{Member, IEEE}, \\ Yanguang Yu, \IEEEmembership{Senior Member, IEEE}, Jun Li, \IEEEmembership{Member, IEEE}, and Ming Jin \IEEEmembership{Senior Member, IEEE}
\thanks{M. Liu, Q. Guo and Y. Yu are with the School of Electrical, Computer and Telecommunications Engineering, University of Wollongong, Wollongong, NSW 2522, Australia (email:mintil@uow.edu.au; qguo@uow.edu.au; yanguang@uow.edu.au).}
\thanks{C. Zeng and J. Li are with the School of Electronic Engineering, Xidian University, Xi’an 710071, China (e-mail: czeng@xidian.edu.cn; junli01@mail.xidian.edu.cn).}
\thanks{M. Jin is with the Faculty of Electrical Engineering and Computer Science, Ningbo University, Ningbo 315211, China (jinming@nbu.edu.cn).}}

\maketitle

\begin{abstract} 
\rev{This work addresses the problem of tracking maneuvering objects with complex motion patterns, a task in which conventional methods often struggle due to their reliance on predefined motion models. We integrate a data-driven liquid neural network (LNN) into the random finite set (RFS) framework, leading to two LNN-RFS filters. By learning continuous-time dynamics directly from data, the LNN enables the filters to adapt to complex, nonlinear motion and achieve accurate tracking of highly maneuvering objects in clutter. This hybrid approach preserves the inherent multi-object tracking strengths of the RFS framework while improving flexibility and robustness. Simulation results on challenging maneuvering scenarios demonstrate substantial gains of the proposed hybrid approach in tracking accuracy.}
\end{abstract}


\IEEEpeerreviewmaketitle

\section{Introduction}

\rrev{Adaptive maneuvering object tracking has emerged as a promising technique for improving the robustness of state estimation across diverse domains~\cite{ref1}, such as autonomous navigation~\cite{ref2} and radar surveillance~\cite{ref3} and underwater vehicle~\cite{ref4}.}
The primary difficulty arises from modeling multiple targets that exhibit nonlinear, abrupt, or highly dynamic motion characteristics ~\cite{ref5}. Traditional filtering techniques often rely on fixed, predefined motion models, which struggle to generalize in complex scenarios~\cite{ref6}.

To handle object appearance and disappearance in a principled probabilistic framework, the random finite set (RFS) methodology has gained significant attention~\cite{ref7,ref8,ref9,ref10,ref11}. \rev{Notable implementations} include the probability hypothesis density (PHD)~\cite{ref8,ref9} and multiple Bernoulli (MeMBer)~\cite{ref10} filters.  \rev{Nevertheless, the performance of these RFS-based methods remains limited by their reliance on manually specified motion models, particularly in tracking objects with highly maneuverable or uncertain dynamics.} 

Recent progress in neural differential equations and continuous-time learning opens new avenues for improving motion adaptability. Notably, liquid neural networks (LNNs) offer a compact and biologically inspired architecture capable of capturing time-varying dynamics via neural ordinary differential equations (ODEs)~\cite{ref12,ref13}. \rev{This property makes LNNs especially well-suited for modeling complex motion dynamics in maneuvering object tracking.} 

\rev{In this work, we propose a hybrid approach for maneuvering object tracking by integrating LNNs into the RFS framework to enhance the tracking adaptability. The resulting hybrid LNN-RFS filters replace handcrafted kinematic assumptions with a learnable continuous-time dynamic representation, enabling effective modeling of complex and nonlinear motion behaviors. By leveraging the expressive power of neural ODE–based architectures within a principled probabilistic framework, the proposed approach enhances robustness in cluttered environments while preserving the multi-object tracking strengths of RFS methods.}

\section{Background}
\subsection{Liquid Neural Networks}
LNNs are continuous-time models designed to capture complex, time-varying dynamics~\cite{ref12, ref13,ref14}. They represent the evolution of internal latent states through differential equations, enabling adaptive temporal resolution and history-dependent behavior.
\rrev{
The latent state $u(t)$ of a liquid time-constant (LTC) network can be determined by solving the follow ODE ~\cite{ref13,ref14}:
\begin{equation}
    \begin{aligned}
        \frac{du(t)}{dt} &= \big[-\eta_\tau + f(u(t), x(t), \theta)\big] \odot \rev{u(t)} \\
        &\quad + A \odot f(u(t), x(t), \theta),
    \end{aligned}
    \label{eq:lnn}
\end{equation}
where $u(t) \in \mathbb{R}^{D}$ is the latent state with $D$ cells at time $t$, 
$\eta_\tau \in \mathbb{R}^{D}$ is a vector of time-constant parameters that dynamically modulates 
the response speed and memory length of the network, $x(t) \in \mathbb{R}^{n_x}$ is the input signal with $n_x$ features, $f(\cdot)$ represents a neural network parameterized by $\theta$, 
the operator $\odot$ denotes the Hadamard (element-wise) product, 
and $A \in \mathbb{R}^{D}$ is a bias vector.
}

\subsection{RFS Filtering}

From a Bayesian filtering perspective, RFS filters model the evolution of a random finite set of object states $x$, denoted as $\mathcal{X} = \{x^{(i)} \mid i = 1, \dots, N\}$ with the cardinality $|\mathcal{X}| = N$ being a random variable. This set is recursively propagated and updated over time within the filtering framework.

\subsubsection{PHD Filter \cite{ref8}}
\rev{The PHD filter propagates the first-order moment density  $D_k(x)$ of an RFS $\mathcal{X}$ of object states through two steps:}

\textit{Prediction:}
\begin{equation}
\begin{split}
D_{k+1|k}(x) = D^B_{k+1}(x) + \int p_S f_{k+1}(x \mid \xi) \times D_{k}(\xi) \, d\xi
\end{split}
\label{eq:phd_pred}
\end{equation}
where $p_S$ is the survival probability \rrev{of object state $\xi$}, $f_{k+1}(x \mid \xi)$ is the probability density function (PDF) of state transition, \rev{and} $D^B_{k+1}(x)$ is the birth intensity.

\textit{Update}: Given a measurement set $Z = \{z^{(1)}, \ldots, z^{(m)}\}$
\begin{align}
    D_{k+1}(x) 
&= \left[1 - p_D(x)\right] D_{k+1|k}(x) + \\
&\quad  \sum_{z \in Z} \frac{p_D(x) g_k(z \mid x) D_{k+1|k}(x)}{\kappa_{k+1}(z) + \int p_D(\xi) g_{k+1}(z \mid \xi) D_{k+1|k}(\xi) \, d\xi}
\label{eq:phd_update}
\end{align}
where $D_{k+1}(x)$ is the posterior intensity, $p_D(x)$ is the detection probability, $g_{k+1}(z \mid x)$ is the measurement likelihood, and $\kappa_{k+1}(z)$ is the clutter intensity.

\subsubsection{MeMber Filter \cite{ref10} }

The MeMber filter propagates a multi-Bernoull RFS by marginalizing data association. The multi-object posterior at time $k+1$ is represented as $\Gamma_{k+1} = \left\{ \left( r_{k+1}^{(i)}, p_{k+1}^{(i)}(x) \right) \right\}_{i=1}^{N_{k+1}}$ with the existence probability  $r_{k+1}^{(i)}$ and the spatial density $p_{k+1}^{(i)}(x)$ of the $i$-th Bernoulli component. \rev{The prediction and update steps are as follows:}
\begin{figure*}[t]
\centerline{\includegraphics[width=0.8\textwidth]{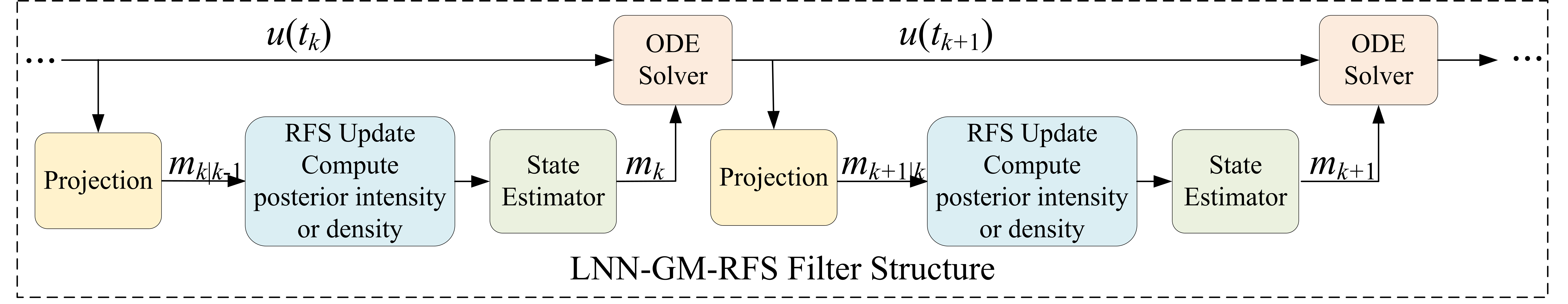}}
\caption{\rev{Structure of the proposed hybrid LNN-RFS filtering.} }
\label{fig:structure}
\end{figure*}

\textit{Prediction:} Each component is predicted as
\begin{align}
    r_{k+1|k}^{(i)} &= p_S \cdot r_{k}^{(i)}, \\
    p_{k+1|k}^{(i)}(x) &= \int f_{k+1}(x|\xi) \, p_{k}^{(i)}(\xi) \, d\xi \label{mb_pre} , 
\end{align}
\rev{New objects} are introduced via a birth model as $\left\{ \left( r_B^{(j)}, p_B^{(j)}(x) \right) \right\}_{j=1}^{N_B}$.

\rrev{\textit{Update:} The update procedure is carried out as described in \cite{ref10}.}

\subsection{\rev{Motivation}}
\rev{While the PHD and MeMBer filters provide principled frameworks for multi-object tracking within the RFS formalism, their performance hinges on the choice of underlying motion models in the prediction step. In practice, conventional models such as constant velocity (CV), constant acceleration (CA), and coordinated turn (CT) rely on fixed-order kinematics and often struggle to accommodate abrupt maneuvers or unknown motion patterns. To address this limitation, we propose hybrid LNN-RFS filtering that integrates LNNs into the RFS framework. By leveraging LNN-based motion prediction, these filters are able to capture complex, time-varying behaviors with improved temporal adaptability, while preserving the principled probabilistic foundations of RFS filtering.} 

\section{\rev{Hybrid LNN-RFS Filtering for Robust Maneuvering Object Tracking}}

\rev{In this section, we first outline the use of LNNs for object motion prediction and describe their integration into RFS filtering framework. This integration enables principled representation of complex and time-varying object dynamics within a Bayesian multi-object tracking setting. Building on this foundation, we then develop two specific filter implementations: the LNN-PHD filter and the LNN-MeMBer filter.}

\subsection{\rev{LNN for Motion Prediction}}
\rev{The LNN is trained to learn object motion dynamics directly from trajectory data. In the radar tracking context, the input to the LNN at time step $k$ is the object’s kinematic state vector $x_{k}$, which typically includes position and velocity components (and can be extended to higher-order motion features if available). The latent state of the LNN at next time instant is projected to the predicted next state $\hat{x}_{k+1|k}$ through a project matrix. The training data consist of state trajectories  
$\{x_0,\, x_1,\, \ldots,\, x_T\}$, either generated from simulated maneuvering object scenarios (e.g., coordinated turns, abrupt accelerations) or collected from  radar measurements.}  

\rev{During training, the network parameters $\theta$ are optimized to minimize the mean squared error (MSE) between the true state $x_{k+1}$ and the predicted state $\hat{x}_{k+1|k}$:  
\begin{align}
\mathcal{L}_{\text{MSE}} = \mathbb{E}\!\left[\|x_{k+1} - \hat{x}_{k+1|k}\|^2\right],
\end{align}
where $\mathbb{E}[\cdot]$ denotes expectation and $\|\cdot\|$ is the Euclidean norm. The parameters are updated using the Adam optimizer~\cite{ref15}
$\theta^{*} = \theta - \eta \cdot \text{Adam} \big(\nabla_{\theta}\mathcal{L}_{\text{MSE}}, \theta\big)$,
with learning rate $\eta$ and gradient $\nabla_{\theta}\mathcal{L}_{\text{MSE}}$ with respect to $\theta$.} 

\rev{Once trained, the LNN replaces handcrafted motion models in the prediction step of RFS filters. In this way, the RFS framework leverages a data-driven, continuous-time dynamic model that can adapt to nonlinear and time-varying object behaviors.} 
\subsection{\rev{Hybrid LNN-RFS Filter Design} }
\rev{The overall structure of the proposed LNN-RFS filter, which governs the object tracking process, is illustrated in Fig.~\ref{fig:structure}. An ODE solver discretizes the continuous interval $[0, T]$ into a sequence of time points $\{t_0, t_1, \ldots, t_K\}$. At each solver step, the latent state is propagated from $t_k$ to $t_{k+1}$ according to the LNN dynamics~\cite{ref13}. The updated latent state is projected onto the predicted object state, which is then passed to the RFS filtering stage. The RFS update computes either the intensity function (PHD filter) or the object state densities (MeMBer filter), after which the estimator outputs the object state. This estimate is fed back as the next input to the ODE solver, forming a hybrid filtering pipeline.} 
\rev{In the following, we elaborate the prediction and update steps of PHD and MeMber filter empowered by the trained LNN.} 

\rrev{Consider zero-mean process noise $v\sim \mathcal{N}(0,Q)$, the state transition model can be described by LNN mapping with network parameters $\theta$ as follows
\begin{equation}
x = \Phi^{\text{LNN}}_\theta(\xi)+v,
\label{eq:delta_transition}
\end{equation}
which determines transition PDF $f_\theta(x|\xi) \sim \mathcal{N}(\Phi^{\text{LNN}}_\theta(\xi),Q)$ with parameterized neural networks $\Phi^{\text{LNN}}_\theta(\cdot)$.} 




\rrev{For nonlinear multiple object model, the intensity function $D(x)$ and the space PDF $p(x)$ can be implemented using either a particle-based (PF) or Gaussian mixture (GM) approach. In this work, the GM formulation is adopted to approximate the object intensity or state density. Subsequently, the (\ref{eq:phd_pred}) and (\ref{mb_pre}) can be modified as}
\begin{align}
D_{k+1|k}(x) &\approx 
    \sum_{j=1}^{J_B} w_B^{(j)} \mathcal{N}(x \mid m_B^{(j)}, Q_B^{(j)}) \notag \\ \nonumber
& + \sum_{j=1}^{J_S} \int p_S(\xi)\, w_S^{(j)} \mathcal{N}(\xi\mid m_k^{(j)}, P_k^{(j)}) \\ 
    & \quad \times \mathcal{N}(x\mid \Phi^{\text{LNN}}_\theta(\xi),Q)\, d\xi,
    \label{eq12}
\end{align} 
Expand $\Phi^{\text{LNN}}_\theta(\xi)$ in a first-order Taylor series around $\xi= m_k$
\begin{equation}
    \Phi^{\text{LNN}}_\theta(\xi)\approx \Phi^{\text{LNN}}_\theta(m_k)+ J_{\Phi}(\xi-m_k),
    \label{eq13}
\end{equation}
\rrev{where Jacobian matrix $J_{\Phi}=\frac{\partial \Phi^{\text{LNN}}_\theta(\xi)}{\partial \xi}\mid _{\xi=m_k}  \in \mathbb{R}^{n_x\times n_x}$}. 

Substitute (\ref{eq13}) into (\ref{eq12}) results in the prediction for survival intensity 
\begin{equation}
     D^S_{k+1|k}(x) \approx \sum_{j=1}^{J_S} p_S\, w_S^{(j)} \mathcal{N}(x\mid m_{k+1}^{(j)}, P_{k+1}^{(j)}),  
\end{equation}
 \rrev{where $J$ and $w$ denote, respectively, the number and the weight of the $j$-th Gaussian component $\mathcal{N}(x \mid m, P)$ in the $D(x)$ or $p(x)$. Each component's predicted mean and covariance matrix are respectively propagated by evaluating the LNN}
\begin{align}
    m_{k+1|k}^{(j)}& = \Phi_\theta^{\text{LNN}}\left(m_{k}^{(j)}\right),\\
    P_{k+1|k}^{(j)}& = J_{\Phi}(m^{(j)}_k)P_{k}^{(j)}J_{\Phi}(m^{(j)}_k)^{\top}+ Q_k.
\end{align}

\rrev{As the analytical derivation of the Jacobian matrix is cumbersome, we employ a forward difference scheme to obtain its numerical approximation
\begin{equation}
    J_{\Phi}(m^{(j)}_{k,l}) \approx \frac{\Phi^{\text{LNN}}_\theta(m^{(j)}_k+he_l)-\Phi^{\text{LNN}}_\theta(m^{(j)}_k)}{h}, \quad l=1,\dots,n_x,
    \label{eq14}
\end{equation}
where $e_l$ is the $l$-th standard basis vector, and $h$ is a small step size, $J_{\Phi}(m^{(j)}_k) = [J_{\Phi}(m^{(j)}_{k,1}),\dots,J_{\Phi}(m^{(j)}_{k,n_x})]$ with $J_{\Phi}(m^{(j)}_{k,l})\in \mathbb{R}^{n_x \times 1}$}.
Similarly, the prediction form of the spatial probability density follows
\begin{equation}
    p_{k+1|k}^{(i)}(x) \approx \sum_{j=1}^{J_i}p_S w^{(i,j)}_B \mathcal{N}(x \mid m_{k+1|k}^{(i,j)}, P_{k+1|k}^{(i,j)}).
\end{equation}
 
 \rrev{Measurement updates follow the standard GM-PHD or GM-MeMber update equations}. The proposed LNN-GM-PHD Filter is summarized in Algorithm \ref{alg:lnn_gm_phd}.
\rrev{ \subsection{The PF implementation}}
\rrev{For the PF implementation, the prediction state is $x_{k+1|k} = \Phi^{\text{LNN}}_\theta(\xi)$, the prediction intensity can be represented a set of weighted particles $\{w^{(j)}_{k+1|k},\xi^{(j)}\}_{j=1}^{J_S}$ for survival objects\cite{ref11} }
\rrev{
\begin{equation}
    D_{k+1|k}(x) \approx \sum_{j=1}^{J_S}w^{(j)}_{k+1|k}\delta(x-x_{k+1|k}^{(j)})
\end{equation}}
\begin{algorithm}[htbp]
\caption{LNN-GM-PHD Filter}
\label{alg:lnn_gm_phd}
\begin{algorithmic}[1]
\REQUIRE Previous intensity $\{w_{k}^{(j)}, m_{k}^{(j)}, P_{k}^{(j)}\}_{j=1}^{J_{k}}$, measurements $Z_{k+1}$, birth components $\{w_{\gamma}^{(j)}, m_{\gamma}^{(j)}, P_{\gamma}^{(j)}\}_{j=1}^{J_{\gamma}}$, LNN model $\Phi_\theta^{\text{LNN}}$
\ENSURE Updated intensity $\{w_{k+1}^{(j)}, m_{k+1}^{(j)}, P_{k+1}^{(j)}\}_{j=1}^{J_{k+1}}$
\STATE \textbf{Prediction Step}
\FOR{each component $j = 1$ to $J_{k}$}
    \STATE Predict mean via LNN: \rrev{$m_{k+1|k}^{(j)} \gets \Phi^{\text{LNN}}_\theta(m_{k}^{(j)})$}
    \STATE Predict covariance: \\ \rrev{$ P_{k+1|k}^{(j)} = J_{\Phi}(m^{(j)}_k)P_{k}^{(j)}J_{\Phi}(m^{(j)}_k)^{\top}+ Q_k$}
    \STATE Predict weight: $w_{k+1|k}^{(j)} \gets p_S \cdot w_{k}^{(j)}$
\ENDFOR
\STATE Add birth components: \\
$\{w_{k+1|k}^{(j)}, m_{k+1|k}^{(j)}, P_{k+1|k}^{(j)}\}_{j=J_{k}+1}^{J_{k}+J_\gamma} \gets \text{birth terms}$
\STATE \textbf{Update Step}
\FOR{each measurement $z \in Z_{k+1}$}
    \FOR{each predicted component $j$}
        \STATE Compute Kalman gain $K^{(j)}$
        \STATE Compute innovation $\nu^{(j)} = z - H m_{k+1|k}^{(j)}$
        \STATE Update mean: $m_{k+1}^{(j,z)} \gets m_{k+1|k}^{(j)} + K^{(j)} \nu^{(j)}$
        \STATE Update covariance: $P_{k+1}^{(j,z)} \gets (I - K^{(j)} H) P_{k+1|k}^{(j)}$
        \STATE Update weight: 
        \[
        w_{k+1}^{(j,z)} \gets \frac{p_D w_{k+1|k}^{(j)} \cdot \mathcal{N}(z \mid H m_{k+1|k}^{(j)}, S^{(i)})}{\kappa(z) + \sum_j p_D w_{k+1|k}^{(j)} \mathcal{N}(z \mid H m_{k+1|k}^{(j)}, S^{(j)})}
        \]
    \ENDFOR
\ENDFOR
\STATE Missed-detection update: \\
$\quad w_{k+1}^{(j,0)} \gets (1 - p_D) \cdot w_{k+1|k}^{(j)}$ \\
$\quad m_{k+1}^{(j,0)} \gets m_{k+1|k}^{(i)}$ \\
$\quad P_{k+1}^{(j,0)} \gets P_{k+1|k}^{(i)}$
\STATE Merge all components:\\
$\{w_{k+1}^{(j)}, m_{k+1}^{(j)}, P_{k+1}^{(j)}\} \gets \text{merge}(\{w_{k+1}^{(j,z)}\}, \{w_{k+1}^{(j,0)}\})$
\STATE Prune low-weight components and limit number to $J_{\max}$
\end{algorithmic}
\end{algorithm}
\section{Simulation Results}
In this section, we evaluate the performance of the proposed LNN-RFS filtering algorithms and compare them with two benchmark approaches for tracking multiple maneuvering \rev{objects}: the interacting multiple model probability hypothesis density (IMM-PHD \cite{ref16}) filter and the interacting multiple model joint probabilistic data association (IMM-JPDA \cite{ref17}) filter. All experiments were performed on a simulation platform equipped with an RTX 4060 laptop GPU and 8 GB of RAM. The implementation was developed using Python 3.9 and PyTorch version 2.5.0.dev20240820+cu121. 
\subsection{Simulation Parameters}
\rrev{A total of 50{,}000 training samples and 3{,}000 test samples were utilized.} The model was trained for 120 epochs with a batch size of 512 and a learning rate of $10^{-3}$
Performance was evaluated using the Generalized Optimal Subpattern Assignment (GOSPA \cite{ref18}) metric, with a cut-off distance of $c =100m$ and an order parameter of 
$p=1$. To assess the LNN model’s ability to capture and predict object maneuvers, we design two scenarios: (i) partial objects multi-mode; (ii) all objects multi-mode.

The object survival probability was set to \(p_S = 0.99\), and the detection probability to \(p_D = 0.95\), the maximum of $J_{max} = 500$ Gaussian components maintained. The pruning threshold was set to \(10^{-5}\), and the component merging threshold to 4. The gating threshold was set to $15$, and the association gating radius was set to $20m$. Track extraction was performed using a threshold of 0.5, and track confirmation was triggered when the track age reached 3 frames. For measurement-driven object birth, the initial weight was set to 0.3, and the standard deviation of the birth and process noise covariance was defined as $Q_\text{B}=\mathrm{diag}([40,\ 40,\ 40,\ 40])$ and \rrev{$Q=\mathrm{diag}([10.0, 10.0, 5.0, 5.0])$}, respectively. The maximum allowable unmatched measurement age was set to 2 frames. \rrev{A step size of $h=10^{-3}$ is adopted for Scenario 1, while a larger step size of $h=10^{-2}$ is used for Scenario 2.} . The total simulation time is set to $K = 100\text{s}$. 
\subsubsection{Partial objects multi-mode scenario}
The sensor surveillance area was set to $3000 m \times 2500m$, the clutter rate is set to $\lambda_c = 40$. There are nine objects with position $[p_x, p_y]$ and velocities $[v_x, v_y]$ in the format $x=[p_x, p_y, v_x,v_y]$, two of which perform multi-mode motion and the rest perform single-mode motion. The IMM model incorporates CV, CT, and CA motion models, with a fixed model transition probability matrix given as follows:
\begin{equation*}
    \pi_{k+1|k}(x|\xi) = \begin{bmatrix}
0.5 & 0.4 & 0.1 \\
0.4 & 0.5 & 0.1 \\
0.1 & 0.4 & 0.5
\end{bmatrix}.
\end{equation*}
\begin{figure}
\centerline{\includegraphics[width=\columnwidth]{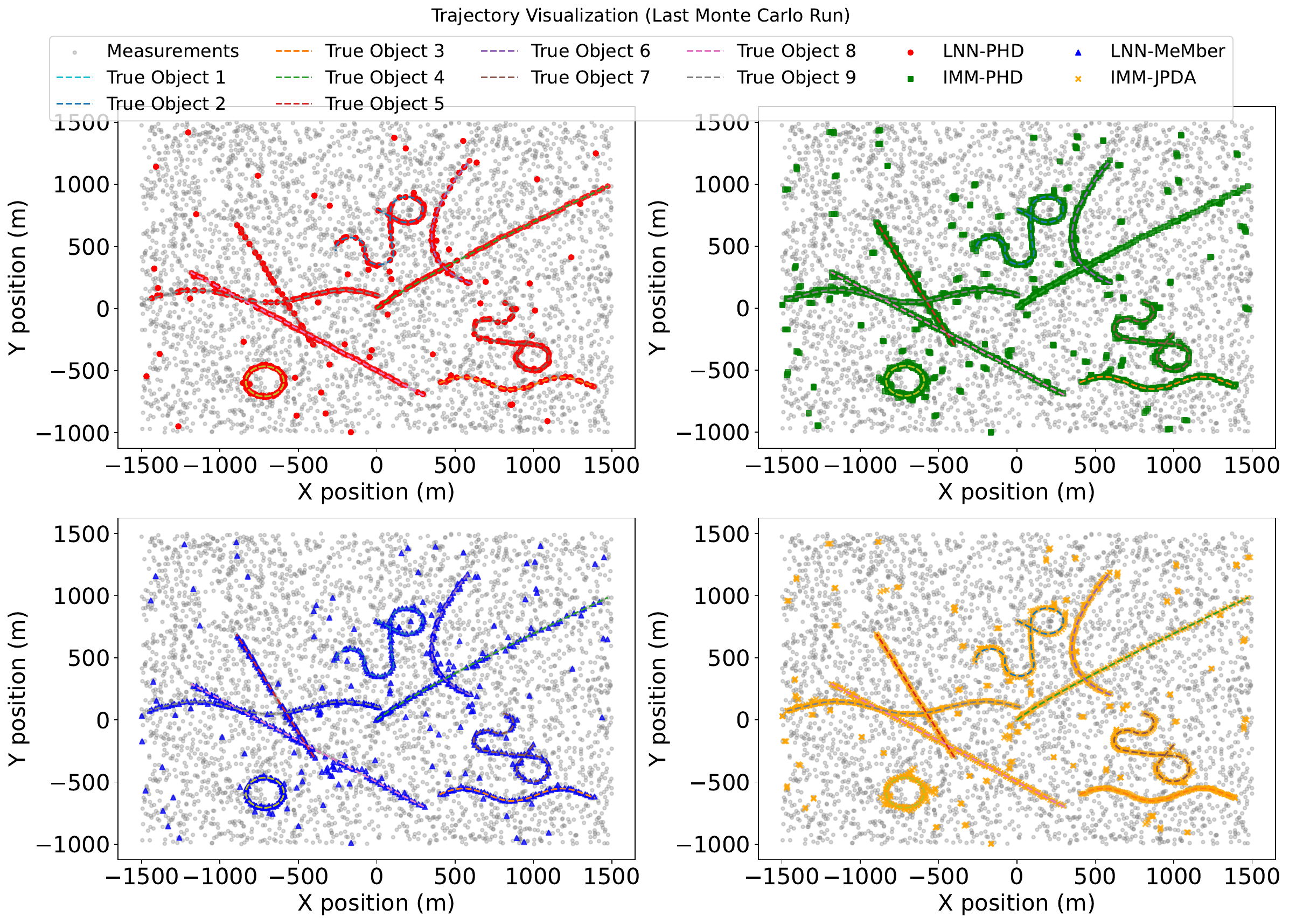}}
\caption{Estimated trajectories of all filters (Scenario 1).}
\label{fig:traj_scenario1}
\end{figure}
\begin{figure}
\centerline{\includegraphics[width=\columnwidth]{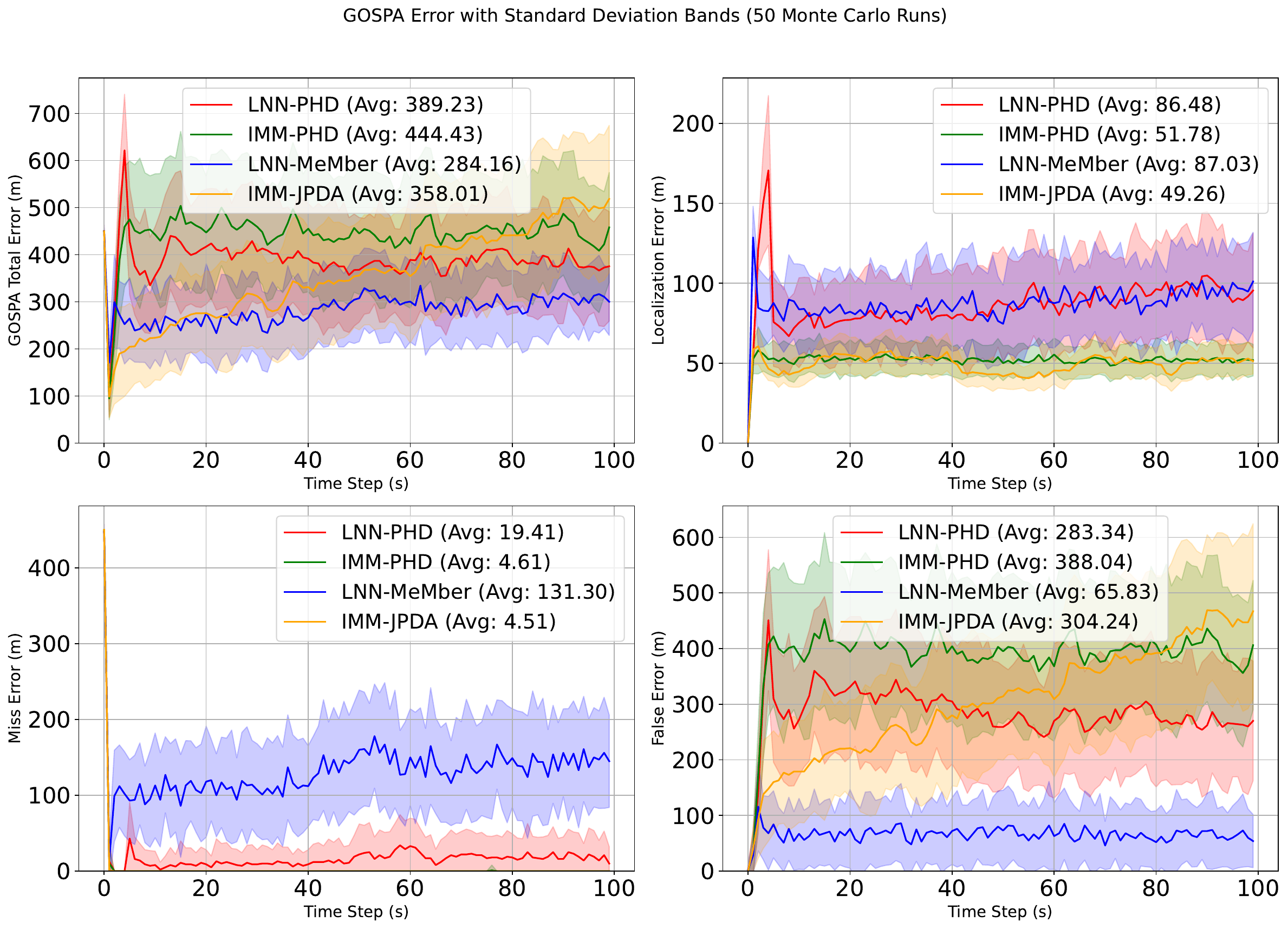}}
\caption{Average GOSPA over all filters vs time (Scenario 1).}
\label{fig:gospa_scenario1}
\end{figure}

The estimated object trajectories and corresponding GOSPA errors are shown in Figs.~\ref{fig:traj_scenario1}--\ref{fig:gospa_scenario1}. In dense clutter scenarios, the proposed LNN-MeMBer filter consistently achieves lower GOSPA errors, owing to its inherent ability to suppress and eliminate components with low existence probabilities, thereby reducing false alarms more effectively than benchmark methods. However, the LNN-MeMBer filter exhibits a higher rate of missed detections, primarily due to its sensitivity to initial state estimates. This underscores the importance of carefully configuring the measurement-driven birth process, particularly the choice of initial birth weight. In contrast, the other filters show lower missed detection rates. Additionally, the IMM-JPDA filter tends to generate a higher number of false alarms over time, which can be attributed to its soft data association strategy being less robust in cluttered environments.
\subsubsection{All objects multi-mode scenario}
\begin{figure}[t]
\centerline{\includegraphics[width=\columnwidth]{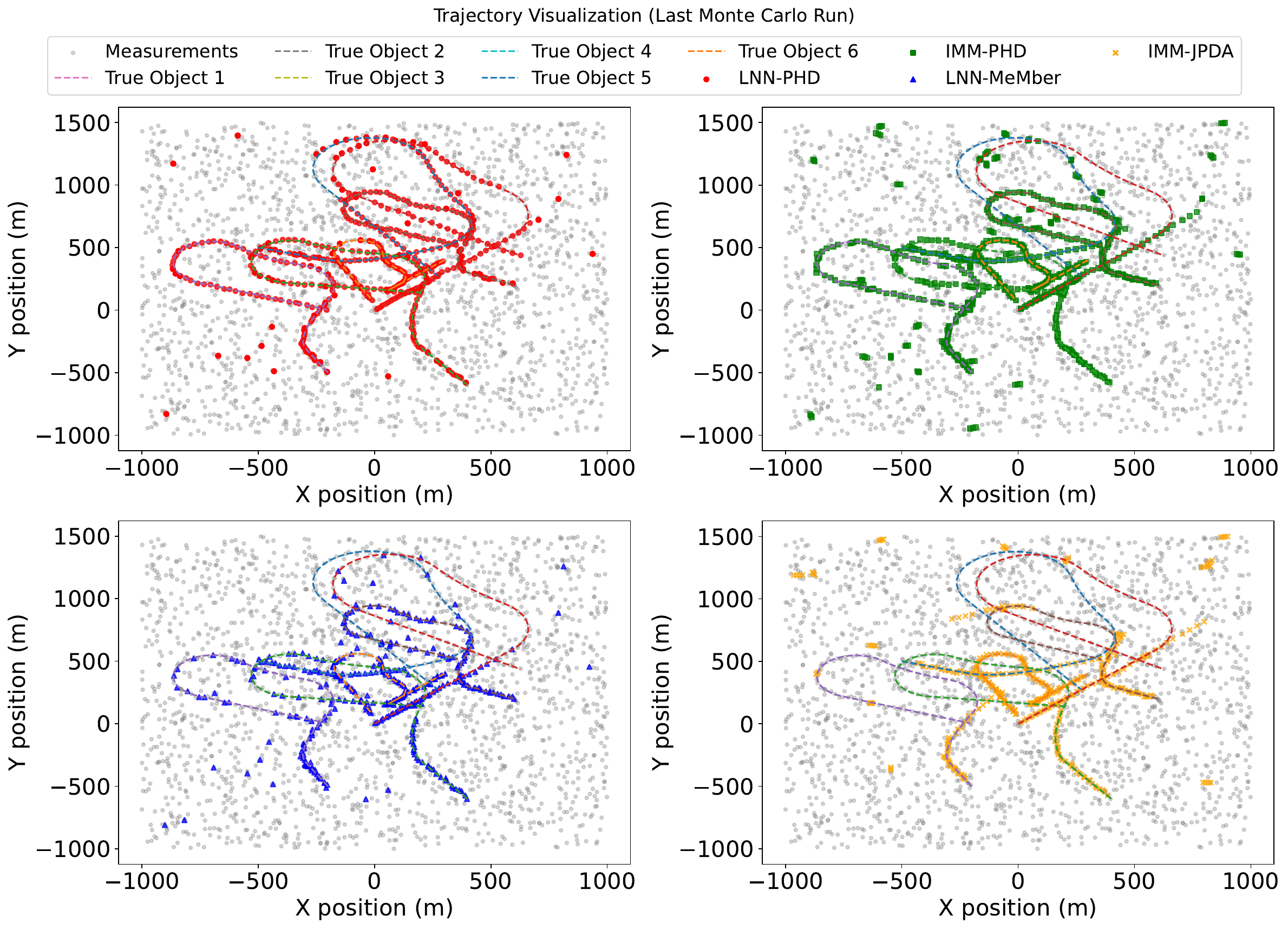}}
\caption{Estimated trajectories of all filters (Scenario 2).}
\label{fig:traj_scenario2}
\end{figure}
\begin{figure}
\centerline{\includegraphics[width=\columnwidth]{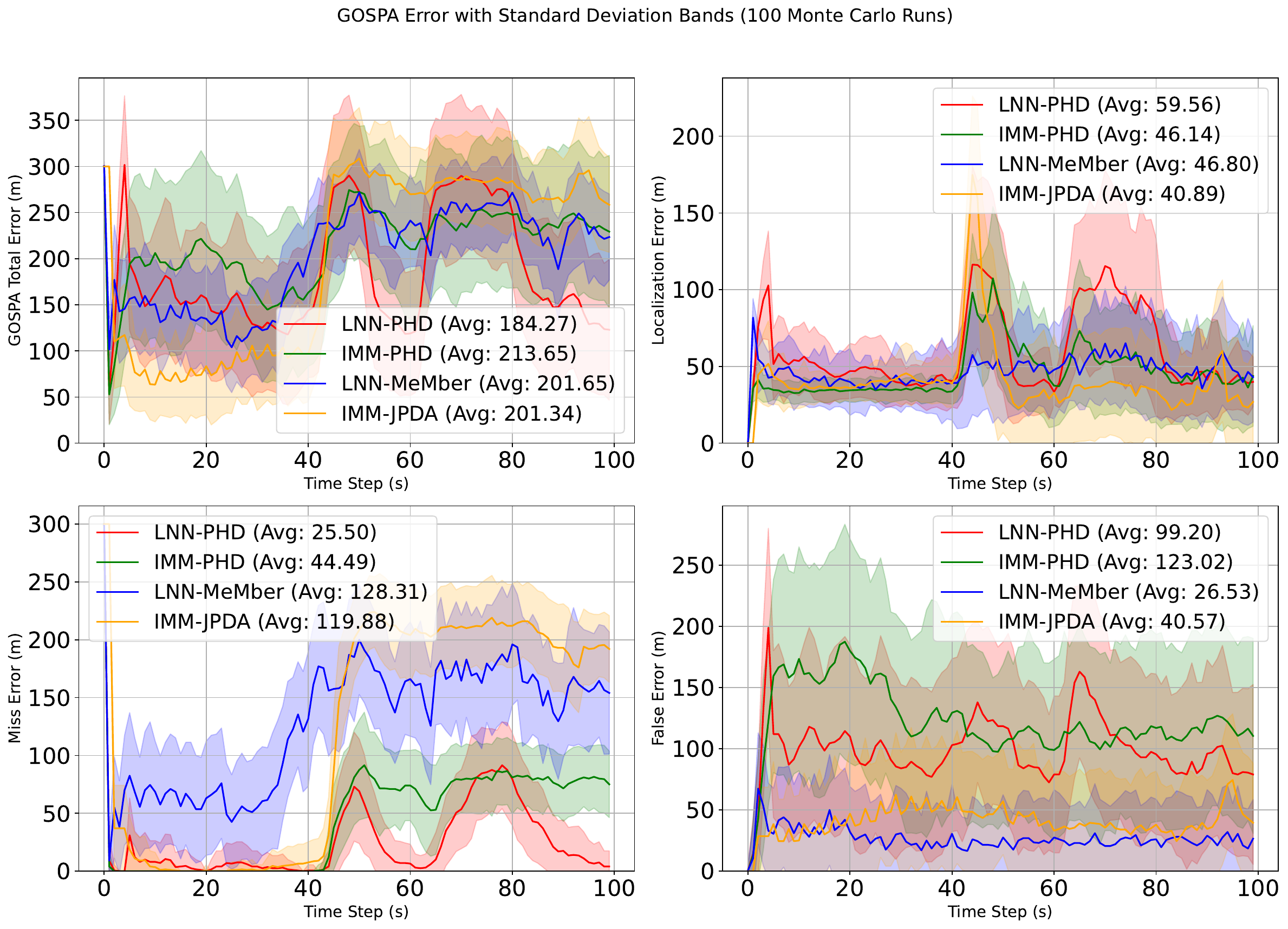}}
\caption{Average GOSPA over all filters vs time (Scenario 2).}
\label{fig:gospa_scenario2}
\end{figure}
The sensor surveillance area was set to $2000 m \times 2500m$, the clutter rate is fixed at $\lambda_c = 20$. There are six highly \rev{maneuvering} objects with the initial positions: $[-500,\ 500,\ 15,\ -5]$, $[300,\ 400,\ -10,\ -10]$, $[400,\ -600,\ -10,\ 15]$, $[0,\ 0,\ 10,\ 10]$, $[-200,\ -500,\ -5,\ 10]$, $[600,\ 200,\ -10,\ 5]$. All objects move with the maneuver modes during different time intervals are defined as follows: 0--20\,s: \texttt{cv}, 20--40\,s: \texttt{ca}, 40--50\,s: \texttt{ct\_left}, 50--60\,s: \texttt{ca}, 60--80\,s: \texttt{ct\_left}, 80--100\,s: \texttt{cv}, in which the turn rate of \texttt{ct}-mode is set to $\omega = 10^\circ$/s, and the acceleration magnitude is fixed at $a_x =a_y = 1\,\text{m/s}^2$. The IMM model incorporates CV, CT, and CA motion models, with a fixed model transition probability matrix given as follows:
\begin{equation*}
    \pi_{k+1|k}(x|\xi) = \begin{bmatrix}
0.4 & 0.3 & 0.3 \\
0.3 & 0.4 & 0.3 \\
0.3 & 0.3 & 0.4
\end{bmatrix}.
\end{equation*}

The tracking performance for all six maneuvering objects is illustrated in Fig.~\ref{fig:traj_scenario2}. It is evident that the proposed LNN-PHD algorithm successfully tracks all objects, whereas the other three methods fail to maintain consistent tracks for some of the highly maneuverable objects. As shown in the GOSPA error results in Fig.~\ref{fig:gospa_scenario2}, all algorithms exhibit similarly low errors during the initial 40 time steps. However, a significant spike in tracking error occurs at \( t = 40\,\text{s} \), corresponding to the transition from \texttt{ca} to \texttt{ct} motion. The magnitude of this error is positively correlated with object maneuverability—higher turn rates lead to larger deviations. Notably, the LNN-PHD filter demonstrates a faster adaptation to motion model transitions and stabilizes more quickly than the other methods. During the transition at \rrev{\( t = 40-60\,\text{s} \) and \( t = 60-80\,\text{s} \) (\texttt{ca} to \texttt{ct}), object 1 experiences a sharp turning maneuver}, only LNN-PHD shows a timely and accurate response, while the other approaches remain largely unresponsive. In the presence of sharp object maneuvers, all methods experience a pronounced and sustained increase in missed detection errors.

\section{Conclusion}

This work presented two neural ODE-based RFS filters for maneuvering object tracking. The LNN-PHD filter enables accurate intensity propagation and robust performance under abrupt maneuvers, consistently outperforming \rev{existing} methods in GOSPA accuracy and clutter resilience. While LNN-MeMBer provides lower false alarms in dense clutter, it is more sensitive to birth parameter settings. Overall, the proposed LNN-RFS approach shows promise for complex object behaviors.

\section*{References}

\def\refname{}

\end{document}